\documentclass[aps,twocolumn,groupedaddress,showpacs,floats,longbibliography,amssymb,prl]{revtex4-1}
\usepackage[dvips]{graphicx}
\usepackage{bm}
\usepackage{color}
\def\he4{$^4$He}
\def\hel3https://www.overleaf.com/project/63cd6fabc5e39bc51b60d244{$^3$He}
\def\Am3{\AA$^{-3}$}
\def\beq{\begin{equation}}
\def\eeq{\end{equation}}

\begin{document}

\noindent {\bf Comment on ``Absence of Off-Diagonal Long-Range Order in hcp $^{4}$He Dislocation Cores"}\vspace{2mm}\\



In a recent Letter~\cite{deKoning2023},  de Koning {\em et al.} report results of first-principle computer simulations of bulk solid ({\em hcp}) $^4$He, in the presence of a single dislocation (various types thereof are considered). The calculation, carried out at zero temperature, shows that the one-body density matrix (OBDM), {\it averaged over the whole system}, decays in the same fashion as in a perfect crystal. This is interpreted as the absence of off-diagonal long-range order, and therefore of superfluidity inside the dislocation core. According to the authors, these results are inconsistent with the superfluid dislocation network scenario \cite{Boninsegni2007} and  invalidate the superclimb mechanism \cite{Soyler2009}, which was further expounded in Refs.~\cite{Pollet2008, Kuklov2022} as the explanation for the unique features of the superflow through solid $^4$He effect \cite{Hallock2008,Hallock2014,Shin2017}.

In this Comment, we contend that the results of de Koning {\it et al.} do not support this conclusion, nor are they inconsistent in any way with the results and predictions of Refs.~\cite{Boninsegni2007,Soyler2009,Kuklov2022,Pollet2008}. 
We explain the origin of the apparent controversy and how to resolve it.
An independent $T=0$ calculation of OBDM {\em inside} the dislocation core would be quite valuable as a stringent test of the predictions made by finite-$T$ worm algorithm Monte Carlo approach Ref.~\cite{Boninsegni2006,Boninsegni2006b} featuring an inherent access to this observable. 
Specifically, the OBDM, defined as
\begin{equation}
\rho ({\bf r}_1,{\bf r}_2) = \langle \hat\Psi^\dagger({\bf r}_2)\hat\Psi({\bf r}_1)\rangle \, ,
\end{equation}
where $\hat\Psi$, $\hat\Psi^\dagger$ are the Bose field operators and $\langle ...\rangle$ stands for the (ground-state) expectation value, is a function of both the relative distance ${\bf r}={\bf r}_2-{\bf r}_1$ {\em and} the center-of-mass position ${\bf R}=({\bf r}_1+{\bf r}_2)/2$.
Superfluidity (or absence thereof) in the dislocation core is revealed
through the slow power-law decay of $\rho$ as a function of $r$ when 
{\it both} ${\bf r}_2$ and ${\bf r}_1$ are located in the core. Otherwise, one ends up probing mainly properties of the insulating  crystal outside the core, and the dislocation signal can easily remain undetectable within the error bars. Ignoring the crucial fact that the dislocation contribution to $\rho({\bf r}_1,{\bf r}_2)$ is highly nonuniform in terms of its dependence on $\bf R$ and highly anisotropic in terms of its dependence on $\bf r$ and, correspondingly, processing the data as if OBDM is a function of the scalar variable $r=|{\bf r}_1-{\bf r}_2|$ leads to an enormous suppression of the dislocation signal---by the factor $\sim L^4$ at $r\sim L$, where $L$ is the linear size of the simulated sample {\color{black} in units of the interparticle distance}. {\color{black} In the calculation of de Koning {\em et al.}, 
$\rho({\bf r}_1,{\bf r}_2)$ is averaged over the entire system
for a given $r$;
the
``bulk condensate density," $n_0$, extracted from such averaged $\rho (r)$
vanishes in the thermodynamic limit faster than $1/L^4$,
making any comparison of their results with those of Refs.~\cite{Boninsegni2007,Soyler2009,Kuklov2022,Pollet2008}, and the ensuing claim of disagreeing physical conclusions,  meaningless: In a system whose linear size $L$  exceeds ten times the interparticle distance, the quantity $n_0$ is  guaranteed to be smaller than that of corresponding bulk superfluid by a factor greater than $10^4$; the data shown in Fig.~2(b) is {\it entirely} consistent with such a behavior ($10^{-2}\times 10^{-4} = 10^{-6}$).}

Furthermore, it needs to be emphasized that, in a finite sample, boundary effects, strain fields, and pressure gradients are unavoidable. All these shift the phase diagram of finite samples. As argued in Ref.~\cite{Kuklov2022}, it is then important to count the local density from the shifted melting density in the simulation cell because both numerically and experimentally the window for superfluidity is very narrow~\cite{Shin2017}. An enhanced local pressure  at the dislocation core in $^4$He suppresses its superfluid response---dramatically or completely.  Since dislocation contribution to bulk-averaged OBDM at interatomic distance  is negligible, the data of  M. de Koning {\em et al.} [Fig.~2(b)] clearly demonstrate that samples with CS and CE dislocations are at elevated bulk pressure---the corresponding OBDMs are suppressed in comparison with the one for the ideal crystal. This renders their ultimate conclusion about the state of dislocation cores---based on comparison with ideal crystals at a lower pressure---unjustified.

Even in the putative absence of local overpressure, the treatment of exchange cycles by De Koning {\em et al.} remains insufficient. They find, by visual inspection of snapshots, no long exchange cycles. 
Here it is important to emphasize  that 
(i) one has to study statistics of exchange cycles in the dislocation cores (individual snapshots are not representative given that even in the liquid at freezing density the condensate fraction is only about 1\%) and
(ii) there is a fundamental difference between measuring exchange cycles and OBDM in PIGS. While the OBDM is a
property of the ground-state wave function, the statistics of exchange cycles is a property of the imaginary-time evolution operator $e^{-\tau H}$ in the path-integral representation. Correspondingly, the projection time $\tau$ for the OBDM can be arbitrarily short---depending on the quality of the trial wave function. But to start seeing long exchange cycles, having an appropriately long $\tau$ is imperative, even when the trial wave function is the exact ground state. Furthermore, in {\it one-dimensional} superfluids, macroscopic exchange cycles appear only when the projection time is macroscopically large, $\tau \propto L$.

In principle, finite-$T$ path integral schemes and $T=0$ projection methods such as PIGS are exact and should give consistent results 
for the same Hamiltonian. It is important to compute the one-dimensional dislocation OBDM and statistics of exchange cycles in the core for identically prepared samples by both methods. Only then one can establish whether the ground state properties 
starting from a (non-orthogonal) trial wave function have been reached.

We acknowledge support from the Natural Sciences and Engineering Research Council of Canada and from the USA National Science Foundation under grants DMR-2032136 and DMR-2032077. L.P. acknowledges support from FP7/ERC Consolidator Grant QSIMCORR, No. 771891.
\vspace{2mm}\\
{\small \noindent  M. Boninsegni,$^1$ A. B. Kuklov,$^2$ L. Pollet$,^3$ N. Prokof'ev,$^4$ and B. V. Svistunov$^4$\\
$^1$Department of Physics, University of Alberta, Edmonton, AB, Canada, T6G 2H5.\\
$^2$Department of Physics and Astronomy, The College of Staten Island, and
the Graduate Center, CUNY, Staten Island, NY 10314.\\
$^3$Arnold Sommerfeld Center for Theoretical Physics, Ludwig-Maximilians Universit\"at, Theresienstrasse 37, 80333 M\"unchen, Germany\\
$^4$Department of Physics,  University of Massachusetts, Amherst, MA 01003.}
\   \bibliography{comment}
\end{document}